# Emergence of π-Magnetism in Fused Aza-Triangulenes: Symmetry and Charge Transfer Effects

*Jan Patrick Calupitan,[1,2,*,‡,†] Alejandro Berdonces-Layunta,[1,2,†] Fernando Aguilar-Galindo,[3] Manuel Vilas-Varela,[4] Diego Peña,[4] David Casanova,[2,5] Martina Corso,[1,2] Dimas G. de Oteyza,[1,2,6,*] Tao Wang,[1,2,*]*

[1] Centro de Física de Materiales (CFM-MPC), CSIC-UPV/EHU, 20018 San Sebastián, Spain

[2] Donostia International Physics Center, 20018 San Sebastián, Spain

[3] Departamento de Química, Universidad Autónoma de Madrid, 28049 Madrid, Spain

[4] Centro Singular de Investigación en Química Biolóxica e Materiais Moleculares (CiQUS) and Departamento de Química Orgánica, Universidade de Santiago de Compostela, 15782 Santiago de Compostela, Spain

[5] Ikerbasque, Basque Foundation for Science, 48009 Bilbao, Spain

[6] Nanomaterials and Nanotechnology Research Center (CINN), CSIC-UNIOVI-PA, 33940 El Entrego, Spain

[‡] Current address : Sorbonne Université, CNRS, Institut Parisien de Chimie Moléculaire, IPCM, F-75005 Paris, France

[†] These authors contributed equally

ABSTRACT. On-surface synthesis has paved the way towards the fabrication and characterization of conjugated carbon-based molecular materials that exhibit π-magnetism such as triangulenes. Aza-triangulene, a nitrogen-substituted derivative, was recently shown to display rich on-surface chemistry, offering an ideal platform to investigate structure-property relations regarding spin-selective charge transfer and magnetic fingerprints. Herein, we study electronic changes upon fusing single molecules into larger dimeric derivatives. We show that the closed-shell structure of aza-triangulene on Ag(111) leads to closed-shell dimers covalently coupled through sterically accessible carbon atoms. Meanwhile, its open-shell structure on Au(111) leads to coupling via atoms displaying high spin density, resulting in symmetric or asymmetric dimer products. Interestingly, whereas all fused products on Au(111) exhibit similar charge transfer properties, only asymmetric dimers show magnetic fingerprints. We rationalize this in terms of molecular bonding structure and π-conjugation: in contrast to the symmetric and highly conjugated dimer, asymmetric dimers display more localized orbitals, which result in a larger Coulomb repulsion and thus promote single electron occupancies with associated spin density and π-magnetism. These results expose a clear relationship between molecular symmetry, charge transfer, and spin states in π-conjugated carbon-based nanostructures.

Mechanistic origins of magnetism in materials can be viewed as breaks in symmetry that allow for unpaired electrons. Though long-time associated with *d*- and *f*-shell electrons of metals, unpaired electrons can also emerge in *p*-shell electrons of carbon-based materials, conferring magnetic properties to π-conjugated structures[1,2] due to: (1) sublattice imbalance[3,4] in π electrons of the graphene hexagonal lattice that introduce a non-zero net spin; (2) topological frustration[5,6] of π electrons, preventing double occupation of the same π-bond; and (3) spin polarization[7–11] of low-energy states driven by Coulomb repulsion or topological phase transitions. Magnetism of carbon nanomaterials is therefore very sensitive to the most trivial changes in chemical structure so that engineering π-magnetism requires precise control over it. Multi-step chemical synthesis can offer such precision, but the intrinsic reactivity of π-radicals renders the the synthesis and characterization by solution chemistry challenging. On-surface synthesis circumvents these problems, simultaneously affording atomic precision, an inert environment, and substrate-stabilization,[12,13] as it has proven effective in fabricating a myriad of π-magnetic molecular materials[1,2] such as triangulenes,[4,14–16] graphene nanoribbons,[7,8] and other graphene nanofragments.[5,9–11,17]

In addition to the intrinsic magnetic properties of such carbon structures, charge transfer between a molecule and a metal surface can also give rise to the generation (or modifications) of π-magnetism.[18,19] Adding or removing one (or more) electron to/from the organic molecule may modify the spin balance. For this to happen, spin-selective charge transfer must occur. That is, the number of α and β spin electrons transferring from (to) the surface is different. Otherwise, if α and β electrons contribute similarly to the charge transfer, the net spin of the molecule will remain unchanged, as happens with conventional charge transfer scenarios resulting in the depopulation of frontier molecular orbitals.[20–25] Although each case has been widely reported, a better understanding of the parameters impacting on the occurrence of magnetism *vs.* conventional orbital depopulation by charge transfer could allow their rational tuning.

Triangulene is the most studied molecule displaying π-magnetism due to sublattice imbalance.[26,27] Although two groups recently succeeded in the solution-phase synthesis of a kinetically-stabilized derivative,[28,29] on-surface techniques have since produced and characterized larger and smaller derivatives,[14–16,30] dimers,[6] trimers,[31] and larger oligomers[32,33]. Inspired by how heteroatom doping enriches the chemistry and physical properties of nanographenes, we recently reported the synthesis and characterization of a nitrogen-doped triangulene, aza-triangulene (**1**, Scheme 1). Different from all-carbon triangulenes, the existence of a nitrogen atom triggers charge transfer between **1** and metal surfaces, showing opposite direction on Au(111) and Ag(111) (Scheme 1).[34] On the latter, **1**$^-$ turned out to display a closed-shell character structure, while on Au(111) **1**$^+$ exhibits an open-shell character (S=1) with spin-density distribution mostly along the edge atoms on the three sides of the molecule.[34] Therefore, **1**, in particular upon fusion into larger derivatives, offers an ideal platform to systematically investigate the parameters influencing the spin relevance of charge transfer to (from) metal surfaces, as associated with magnetic properties.

In this work, we fabricate various extended triangulene structures by fusion of **1** and investigate their tunable magnetic properties. By means of high-resolution bond-resolving scanning tunneling microscopy (BR-STM) with a CO-functionalized STM tip,[35,36] we show that on Ag(111) the coupling products are random, mostly limited by steric effects, and present closed-shell character. In contrast, on Au(111), a preferential formation of fused dimeric products was observed resulting from straight and staggered alignment of reactive carbon sites (those with highest spin density), producing symmetric and asymmetric structures, respectively. Scanning tunneling spectroscopy (STS) characterization of electronic structure, coupled with density functional theory (DFT) calculations, reveal that although all dimers display comparable charge transfer properties to Au(111), only asymmetric fusion products show magnetic fingerprints. We link the appearance of magnetic fingerprints in asymmetric products to the disruption of π-conjugation patterns, resulting in more localized frontier orbitals and weaker orbital hybridizations. This behavior is associated with a large Coulomb repulsion that leads to spin-selective charge transfer[18,19,37] from the asymmetric dimers to Au(111), forming singly occupied/unoccupied molecular orbitals (SOMO/SUMO), in contrast with conventional HOMO depopulation[20,24,25] in the symmetric dimer. Further breaking the symmetry of the dimers by considering the fusion of non-equivalent units resulted in a high spin S=1 system. This work reveals how molecular symmetry and charge transfer synergistically influence spin states of carbon materials.

We previously reported the on-surface synthesis and characterization of aza-triangulene (**1**, Scheme 1) on Ag(111) and Au(111).[34] Briefly, a ketone-functionalized precursor was deposited on a metal surface, reduced with atomic hydrogen, and then annealed to 250 ºC, which removed the ketones and left on the surface aza-triangulene with hydrogenated edge atoms. The additional hydrogen atoms present in $sp^3$-hybridized carbon atoms could be removed by the scanning probe tip or by further annealing at higher temperatures, which triggered the formation of fused products on both substrates, with lower thresholds observed on Au(111) (~300 ºC) than on Ag(111) (~350 ºC).[34]

Large-scale STM images on Ag(111) show that products do not follow a regular fusion pattern and no systematically repeating structures were found (Figure 1a). BR-STM images of planar and thus better identifiable structures (Figure 1a, white and pink arrows) revealed molecules resulting from covalent bonding of aza-triangulene units on their vertices. Coupling motifs involving the vertices relate to minimal steric hindrance owing to the available space around which other units could approach independently of their relative alignment (Figure 1b, Figure S1). We note that these products did not show any spectroscopic fingerprint in differential conductance (dI/dV) spectra typically associated with magnetic properties, *e.g.*, zero-bias Kondo resonances or inelastic spin-flip excitation steps (Figure 1c and Figure S1). This behavior points to closed-shell fused products, consistent with the closed-shell structure of the coupling unit **1**⁻.

The situation changes on Au(111), where charge transfer in the opposite direction results in **1$^+$** with degenerate singly (un)occupied orbitals.[34] Spin-density is directly related to an enhanced reactivity in carbon nanostructures,[1,38,39] revealing the most reactive carbon atoms in open-shell structures. The calculated spin density of **1$^+$** (Figure S2) showed that the most reactive sites are the carbon atoms on the edges of this triangular molecule (Scheme 1, carbon atoms in green). This is experimentally confirmed when aza-triangulene units fuse on Au(111). Although large-scale STM images (Figure 2a and S3) show various products, a finite number of systematically repeating structures indicate coupling through preferential carbon sites. BR-STM images revealed the most common dimeric product **2** (Figure 2a; blue arrow), formed by coupling the three reactive carbon atoms of the zigzag edge (Figure 2b; green circles) of two symmetrically aligned monomeric units. This structure accounting for more than half of the fused products is a consequence of the electronic structure of **1$^+$**, since approaching three pairs of neighboring carbon atoms should have been sterically prevented if not for the thermodynamic gain from quenching reactive sites of the monomer, which following the Bell-Evans-Polanyi principle also lowers the kinetic barrier.[40,41]

Other main products appear on large-scale STM images with rhombic (Figure 2a; yellow arrow) and trapezoidal (Figure 2a; green arrow) shapes. BR-STM images of the rhombic species revealed either of two structures **3** (Figure 2c) and **3'** (Figure 2d), the latter of which can be thought of as a rearrangement of the former. Meanwhile, BR-STM images of the trapezoidal species revealed any of the three structures **4**, **4'** and **4''** (Figure 2e-g). **4'** can be thought of as a rearrangement product of **4**, while **4''** results from the oxidative loss of a carbon atom from **4**. We note that rearrangement reactions[42–44] and oxidative carbon loss[45] have been previously observed in the on-surface synthesis of other carbon-based nanostructures. For practical reasons, we used large-scale STM images to report yields of formation collectively. Since **3** and **3'** differ only in the position of the five-membered ring, they could be distinguished only by BR-STM imaging with a CO-functionalized tip. Similarly, **4**, **4'**, and **4''** show almost identical shapes in large-scale STM images. Additionally, BR-STM could reveal hydrogenated derivatives, indistinguishable from the fully π-conjugated products without CO-functionalization (the former requiring tip-induced deprotonation to transform into the latter, as illustrated for **4** in Figure S3). For these reasons, we grouped the products into rhombic (**3**, **3'**) and trapezoidal (**4**, **4'**, and **4''**) species in the statistical analysis, constituting almost a quarter and a sixth of the fused products respectively (Figure 2h).

We interpret such distribution by considering how monomeric units fused to form **3** and **4**. At first glance, **3** and **4** result from a staggered alignment of two monomeric units (Figure 2c and 2e), effectively departing from the symmetric alignment of the monomeric units in **2** (Figure 2b). The formation of one more covalent bond to produce **3** makes it a more exothermic product than **4**, justifying the higher yield of the rhombic than trapezoidal structures. Looking closer at **3**, two of the bonds are through pairs of reactive sites of the monomer (Figure 2c, green lines), while the other (red line) is between a reactive carbon and the vertex of the other. For **4**, two bonds are formed: one is through a pair of reactive carbon atoms (Figure 2d, green line) and another (red line) is through a reactive carbon and the vertex of another molecule. Thus, the yields of fused products (**2** > **3** > **4**) is dictated by the number of effectively "quenched" high-energy reactive carbon atoms bonded to those of another monomer. From these pictures it follows that one or two reactive sites of a monomeric unit are not bonded to a reactive carbon/s of another unit (Figure 2c, 2e, red circles), so that resonance structures of **3** and **4** may be written (Figure S5) to harbor π-radicals (*i.e.*, explicitly showing open-shell structures). Nevertheless, this open-shell structure can be avoided by drawing **3**, **4**, and their derivatives with a zwitterionic pyridinium moiety, as proposed previously for the spin-lowering Jahn-Teller distortion[46] of **1**.[34] Moreover, DFT calculations supported a closed-shell ground state for **3** and **4** (see details in Figure S5).

Diferential conductance spectra of the symmetric dimer product **2** revealed resonances at −1.25 V, 5 mV, 460 mV, and 860 mV (Figure 3a-b). Figure 3b shows a short-range dI/dV spectrum around the Fermi level. Differential conductance (dI/dV) maps obtained with a CO-functionalized tip at these energies (Figure 3c) show qualitative agreement with DFT-calculated density of states (DOS) of the HOMO−1, HOMO, LUMO and LUMO+1 (Figure 3d, the orbital wavefunctions are pictured in Figure S6). Nevertheless, there are apparent discrepancies between the calculated DOS and experimental dI/dV maps, mainly due to the $p_xp_y$-wave character of the CO-functionalized tip used for the mapping of the orbitals.[47,48] A better fit is thus obtained with simulated dI/dV maps considering tunneling through the p-orbitals of the CO-functionalized tip (Figure 3e).[47,48] The resonance corresponding to the HOMO being centered slightly above the Fermi level (Figure 3b) is indicative of its partial depopulation, *i.e.*, charge transfer from **2** to the Au(111) substrate. This agrees well with periodic DFT calculations that predict a charge transfer from **2** to the Au(111) substrate of approximately 1.4 electrons (see details of computational methods in the SI). Henceforth we write **2**$^+$ to emphasize the charge transfer to the surface and the cationic molecular nature.

Relaxation of the asymmetric structures **3** and **4** on Au(111) similarly predicts a charge transfer of 1.11 and 1.15 electrons respectively (Table 1). Indeed, in the same way as it happens with **1**,[34] the calculated oxidation potentials of **2**, **3**, and **4** were all in the same range (5.0-5.2 eV) and lower than the work function of the Au(111) substrate (~5.3 eV). However, in contrast to the closed-shell product **2**$^+$, asymmetric products **3**$^+$ and **4**$^+$ clearly manifest magnetic fingerprints (in particular a zero-bias Kondo resonance) associated with singly occupied molecular orbitals that result from their cationic nature (Figure 3f-j).

Figure 3f shows a representative dI/dV spectrum acquired on **3** while Figure 3h shows conductance maps of the various resonances, in particular at the energies marked with red lines. Whereas the experimental findings cannot be reproduced with calculations of neutral **3**, a reasonable agreement is readily achieved with DFT-calculated DOS of **3**$^+$ in the gas phase (Figure 3i, see Figure S7a for the wavefunctions of the associated states). We identify the peak at −300 mV with a convolution of the HOMO and SOMO; the 175 mV peak with the SUMO; and the 580 mV peak with the LUMO.

Additionally, the spatial distribution of the Kondo resonance (Figure 3h), which can be fitted by a Frota function[49] with a full width at half maximum (FWHM) of 8.6 mV (Figure S8), is consistent with the SOMO and also matches the calculated spin density of $3^+$ with S=1/2 (Figure 3i). Discrepancies between the dI/dV maps and calculated DOS could be due not only to tunneling via the CO tip (as in $2^+$),[47,48] but also to the different molecular distortions between the structure in gas phase and that adsorbed on Au(111) (see Figure S7 for details). Simulated dI/dV maps that match the experimental data (Figure 3j) were generated considering tunneling between the CO functionalized-tip and the relaxed structure of $3^+$ adsorbed on Au(111) (Figure S7 and computational details in the SI).

In a similar manner as **3**, dimer **4** presents dI/dV spectra and maps that match the electronic structure of its cation $4^+$ with a S=1/2 state (Figure S9). We note that a tip-induced dehydrogenation experiment (See Figure S3 and section on High-Spin System) provides further proof of the charge transfer.[17,34,50] These findings on the different dimers are similar to results on monomer **1**, in which the removal of one electron was necessary to reproduce electronic structures measured on Au(111).[34] The low barrier to charge transfer for **1** was concurred recently by studies illustrating the ease of oxidation of derivatives synthesized in the bulk phase.[51,52] We further note the quantitatively comparable charge transfer in monomers and dimers in spite of their varying size. It is also worth contrasting our findings with the lack of charge transfer in all-carbon triangulene[4] and its dimer derivative[17] as a clear example of how nitrogen doping enriches the chemistry of graphenic nanostructures.[24,25,53–55]

Symmetric ($2^+$) and asymmetric ($3^+$, $4^+$) dimer products manifest different magnetic behavior on Au(111). Whereas in $2^+$ charge transfer induces the partial depopulation of the HOMO, in $3^+$ and $4^+$ the electron transfer results in the appearance of SOMO/SUMO and an associated Kondo resonance as typical magnetic fingerprints. Indeed, as shown in Table 1, the calculated magnetization was found to be almost zero (0.002) for $2^+$, but close to one for $3^+$ and $4^+$ (0.77 and 0.95, respectively). $2^+$ does not present magnetic fingerprints because of the equal contributions of α and β electrons to charge transfer, while **3** and **4** essentially transfer a β electron to the Au(111) surface and turn $3^+$ and $4^+$ into open-shell radicals.

Charge transfer-related magnetism is promoted by: (i) transfer of an electron (as close as possible to unity); (ii) a narrow state width; and (iii) large Coulomb repulsion.[1,56] All three points disfavor **2$^+$**: (i) it departs most from a pure single electron transfer scenario (1.4 electrons); (ii) its planar adsorption configuration promotes a better coupling with the substrate, resulting in a stronger hybridization and thereby in a larger state width than for the non-planar **3$^+$** or **4$^+$**; and (iii) Coulomb repulsion is much lower, as evidenced by gas-phase DFT calculations on cationic radical **2$^+$** (Figure S10), which result in a SOMO-SUMO gap of 1 eV lower than that for **3$^+$** (Figure S7a). The lower Coulombic repulsion is due to a larger extent of π-conjugation in **2$^+$**,[1,8,39,57] tied to its symmetric structure.[58] The HOMO of **2** is globally delocalized and distributed uniformly on the molecule (Figure S5). It is worth noting that it is not symmetry *per se*, but the undisrupted π-conjugation that ultimately prevents **2$^+$** from having magnetization. This is in contrast with the asymmetric bonding configuration of **3** and **4**, bearing five-membered rings as fusing motif that disrupt π-conjugation of the hexagonal lattice. Non-benzenoid structures disrupt π-conjugation in a way that carbon atoms from the same sublattice form a π-bond, leading to unfavorable spin frustration.[59] Due to this, the HOMO of the asymmetric products tends to localize on one subunit. Electrostatic forces scaling inversely with the distance, more localized states in the asymmetric dimers display stronger Coulomb repulsion[1,60,61] leading to spin-selective charge transfer. In this case, the depopulated HOMO splits into SOMO and SUMO gapped by the Coulomb energy, in order to lower the total energy of the system.

While **3** and **4** break the symmetric alignment of **2**, they remain fused dimeric pairs of equivalent monomeric units. Breaking the symmetry even further in **4''** by coupling different monomeric units (Figure 2g) allows generating a high-spin S=1 system. **4''** can be found as a doubly-hydrogenated product **4''-2H**, with BR-STM unambiguously revealing the hydrogenated rings by their larger size and characteristic contrast (Figure 4a). In spite of the odd number of $sp^2$ carbons (and therefore odd number of π-electrons in its neutral state) **4''-2H** does not display a Kondo resonance (Figure 4b). However, the molecule displays a strong and narrow Kondo resonance peak (FWHM=9.1 mV, Figure S8) typical of S=1/2 systems after tip-induced dehydrogenation to **4''-H** (Figure 4c), which effectively adds one π-electron to the structure. As in previously discussed molecules, including monomer **1**, experiment and theory could only be reconciled by considering the molecules in their cationic states. Kondo resonance maps obtained at U=5mV match well the calculated spin density of the cationic **4''-H**$^+$ with S=1/2 (Figure 4d). Upon further conversion to **4''** (Figure 4e), *i.e.*, adding another π-electron, the spectrum transforms to a much weaker and wider peak (FWHM=15.0 mV, Figure S8), typical of S=1 systems.[17,34,60,62] The Kondo maps are consistent with the calculated spin density of **4''**$^+$ with S=1 ground state (Figure 4f). In addition, dI/dV maps of the molecular resonances match DFT calculations on the spin-triplet (S=1) cationic **4''**$^+$ species, confirming the presence of two pairs of SOMOs/SUMOs (Figure S11). *A priori* this seems surprising, as one would expect that neutral **4''** (with its odd number of electrons and S=1/2) would transform to a zero-spin ground state upon charge transfer to Au(111) in **4''**$^+$. This naively unexpected scenario results from the unpaired electron in **4''** being in an orbital (SOMO) of lower energy than the HOMO (See Figure S12). Thus, upon charge transfer the substrate abstracts an electron from the HOMO (Figure S11), resulting in **4''**$^+$ with two unpaired π-electrons (S=1). Here the completely broken molecular symmetry directs the spin-selective charge transfer and leads to the formation of a high-spin ferromagnetic system.

In summary, we relate charge transfer and symmetry considerations with the chemical and magnetic properties of aza-triangulene monomers and dimers. Starting with the dimerization process, we show how the molecule's reactivity is determined by their spin density (or the absence thereof). In its negatively charged and closed-shell state (as occurs on Ag(111)), the dimerization processes require a higher activation temperature and result in random products, mainly limited by steric effects. In contrast, positively charged aza-triangulene with a S=1 ground state (as occurs on Au(111)), displays an enhanced reactivity, with a lower thermal activation threshold and driven by the spin density, consequently with a more limited number of products.

All the probed aza-triangulene dimer products on Au(111) reveal a comparable molecule-to-substrate electron transfer, but with disparate effect on the molecule's magnetism: whereas the symmetric dimer product displays a closed-shell structure, the asymmetric dimers unambiguously display magnetic fingerprints. The symmetric dimer $2^+$ is fully planar and has a well-extended π-conjugation. The former promotes the electronic coupling with the metallic substrate (leading to wider orbitals) and the latter promotes a reduced Coulomb repulsion, two factors that disfavour the emergence of magnetism. In contrast, the non-planarity of the asymmetric products reduces their hybridization with the substrate. Besides, the asymmetric bonding configurations disrupt the π-conjugation, causing the electrons to localize and thus suffer stronger Coulomb repulsion. As a result, it becomes favourable for the molecules to adopt single electron occupancies on particular orbitals, with its associated magnetism. We also show that further symmetry breaking by coupling two disparate monomers could lead to high-spin systems. Altogether, this study shows how charge transfer effects, combined with symmetry considerations, can affect the future design of π-magnetic nanostructures.

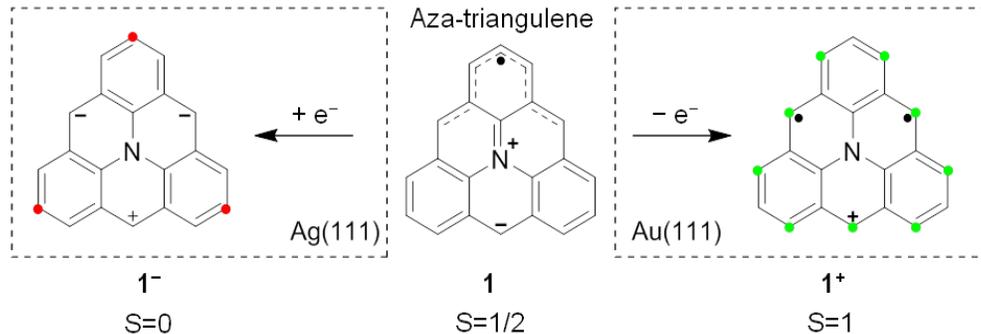

**Scheme 1**. Structure of aza-triangulene (**1**), and its charge transfer properties on Ag(111) and Au(111). On Ag(111), anionic **1⁻** has a closed-shell structure while on Au(111), **1⁺** has an open-shell structure (S=1). The least sterically hindered sites on the **1⁻** are marked in red. The most reactive carbon sites on **1⁺** as determined by its spin density are indicated in green.

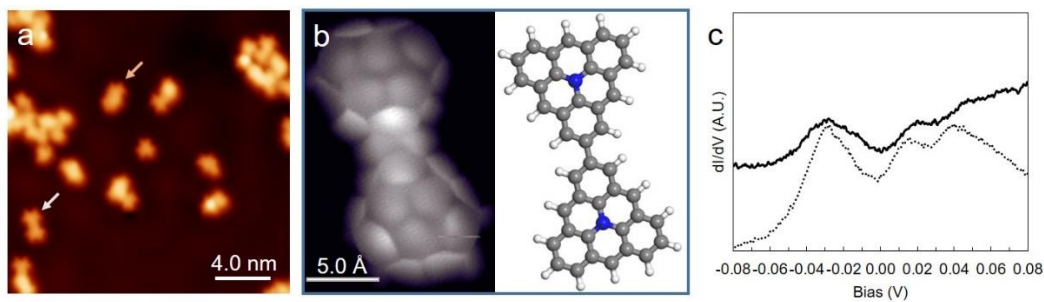

**Figure 1.** (a) Large-scale STM image of random fusion products on Ag(111) (U=−500 mV, I=−100 pA). Arrows point to flat dimeric products resulting to fusion of two aza-triangulene molecules (**1⁻**). (b) High-resolution bond-resolving STM image of the product pointed by the white arrow in (a), together with its corresponding structural model. (c) dI/dV spectra taken on the edge of dimer shown in (b) (solid line) and on bare Ag(111) (dotted line), showing a lack of signals (other than the Ag(111) surface state) that could be attributed to magnetic fingerprints.

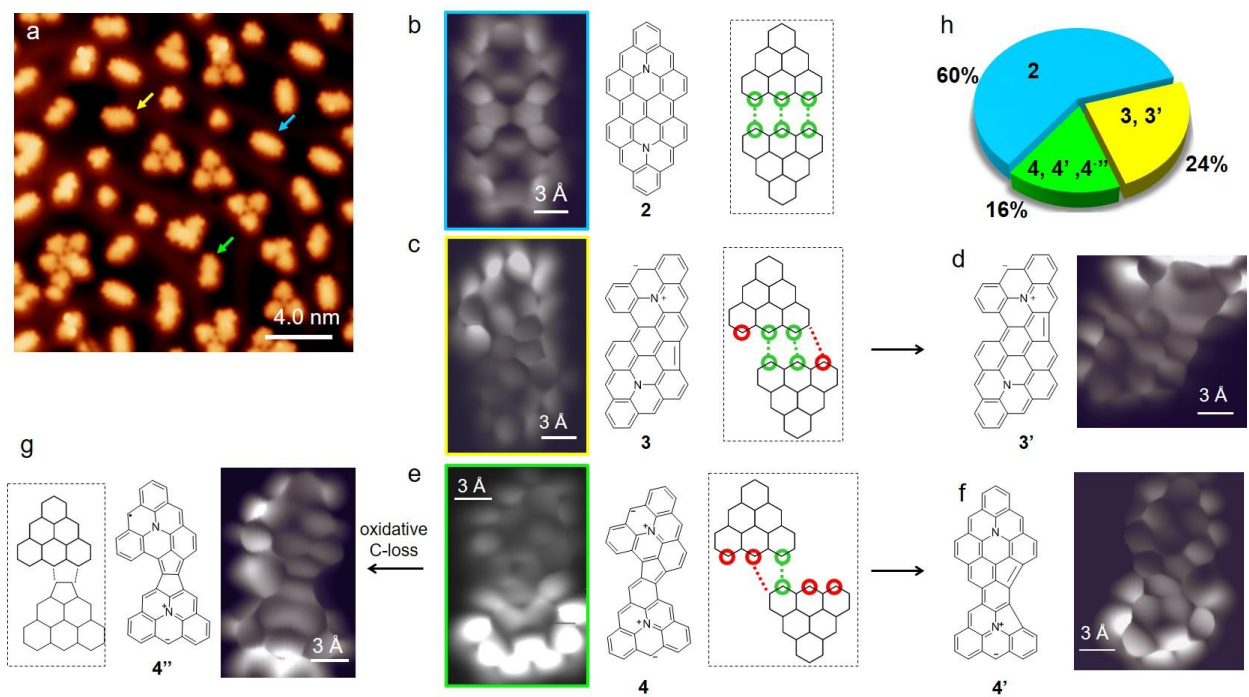

**Figure 2.** (a) Large-scale STM image of products on Au(111) (U=−500 mV, I=100pA). (b-g) Structure and high-resolution bond-resolving STM images with CO-functionalized tip. Blue, yellow, and green arrows point to the rectangular, rhombic, and trapezoidal structures respectively. (b-g) Scheme of bonding alignment, BR-STM images, and chemical structures of fused products. (h) Distribution of products from a statistical analysis of more than 100 fused products.

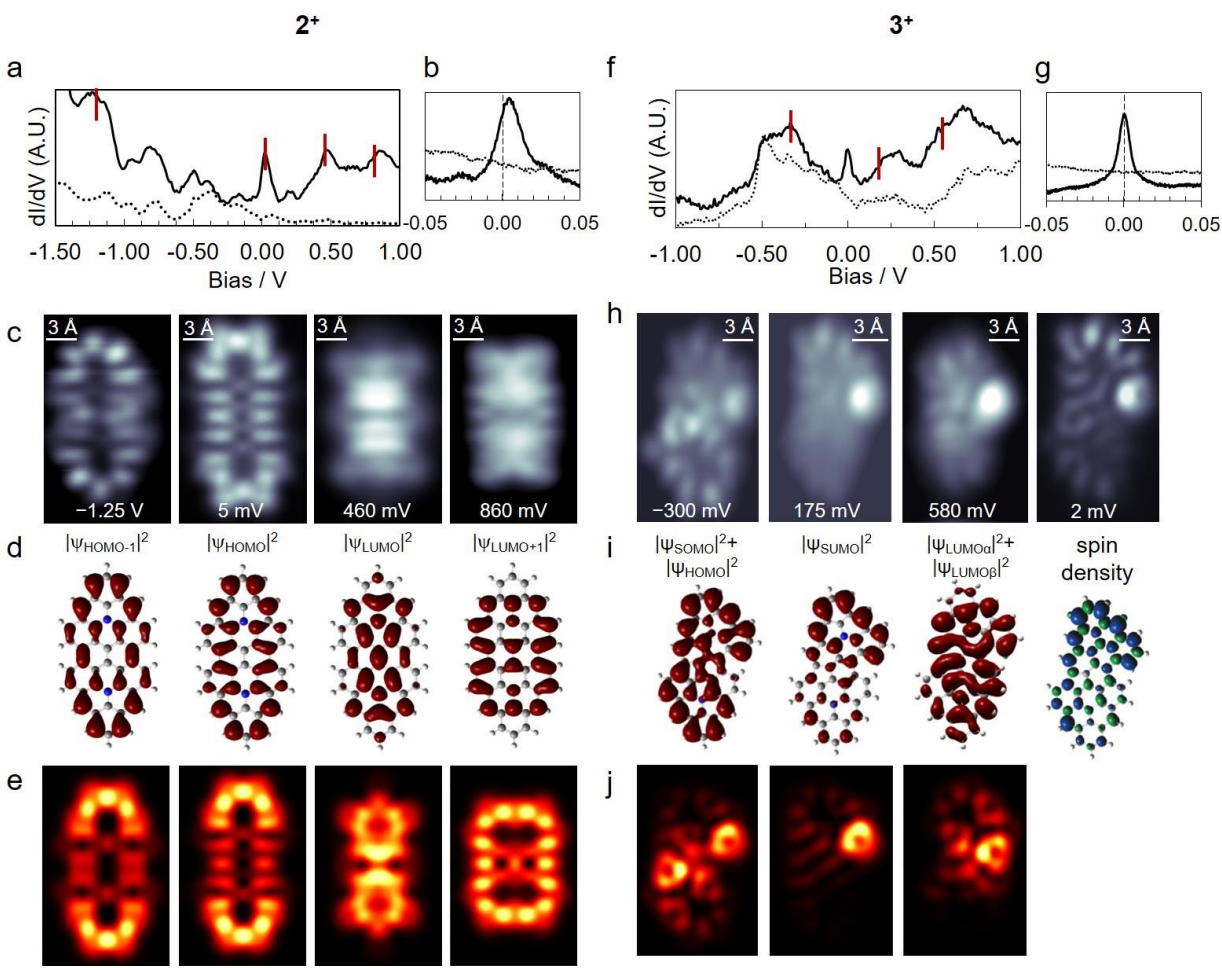

**Figure 3**. Long-range (a,f) and low-energy (b,g) conductance spectra taken on Au(111) (dotted lines) and on dimers **2⁺** and **3⁺** (solid lines); (c,h) dI/dV maps at the designated energies. (d,i) DFT-calculated DOS of dimer **2** and **3⁺** in the gas phase (panel i includes spin density for **3⁺**). (e,j) Simulated dI/dV maps for the orbitals of **2⁺** and **3⁺** listed and shown in panels (c) and (d). The simulations consider a CO-functionalized probe (see details in the SI).

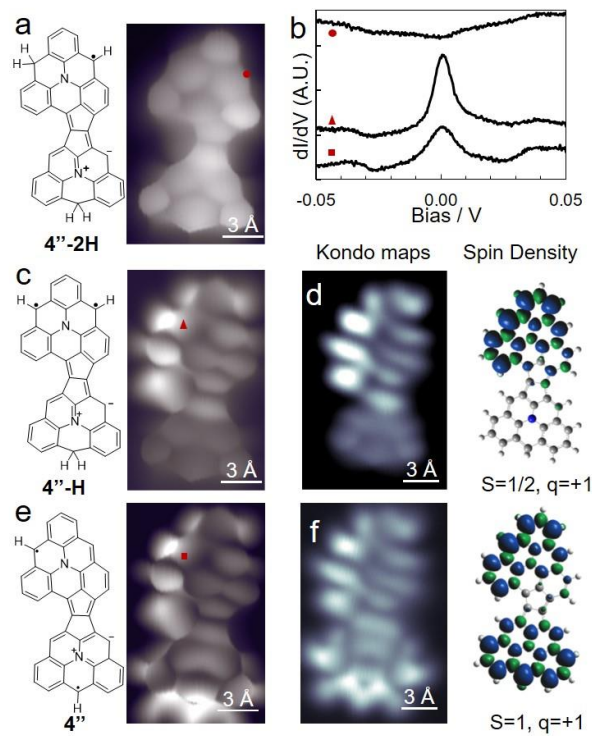

**Figure 4.** (a) Structure and BR-STM image of doubly hydrogenated dimer **4''** (**4''-2H**). (b) Low-energy spectra of **4''-2H**, **4''-H**, **4''** on Au(111). (c,e) Structure and BR-STM image of (c) **4''-H** and (e) **4'''**. (d,f) Constant height STM images acquired at 5 mV corresponding to Kondo maps and calculated spin density (on the right) of (d) **4''-H$^+$** and (f) **4''$^+$**. (a,c,e) Only hydrogen atoms bonded to $sp^3$ carbons or to $sp^2$ radical sites are explicitly shown in the resonant structure.

Table 1. Contribution of α and β electrons to the charge transfer from molecule to Au(111)

| Molecule | Charge Transfer α+β | Magnetization −(α−β) | α contribution | β contribution | % β / % α |
|---|---|---|---|---|---|
| **2** | 1.4 | 0.002 | 0.699 | 0.701 | 50 / 50 |
| **3** | 1.11 | 0.77 | 0.17 | 0.94 | 85 / 15 |
| **4** | 1.15 | 0.95 | 0.10 | 1.05 | 91 / 9 |

## ASSOCIATED CONTENT

**Supporting Information**. Supplementary STM images, DFT calculations, and explanatory notes are available.

The following files are available free of charge.

## AUTHOR INFORMATION

**Corresponding Author**

*jan.calupitan@sorbonne-universite.fr; *d.g.oteyza@cinn.es; *taowang@dipc.org

**Author Contributions**

† These authors contributed equally. The manuscript was written through contributions of all authors. All authors have given approval to the final version of the manuscript.

## ACKNOWLEDGMENT

This work was funded by the Spanish MCIN/AEI/ 10.13039/501100011033 (PID2019-107338RB-C63, TED2021-132388B-C43, PID2019-109555GB-I00), the European Union's Horizon 2020 research and innovation program (Marie Skłodowska-Curie Actions Individual Fellowship no. 101022150), the Basque Government (IT1591-22). F.A.-G. thanks Ministerio de Universidades, Plan de Recuperación, Transformación y Resiliencia and UAM for the Programa de Recualificacion (CA5/RSUE/2022-00234).

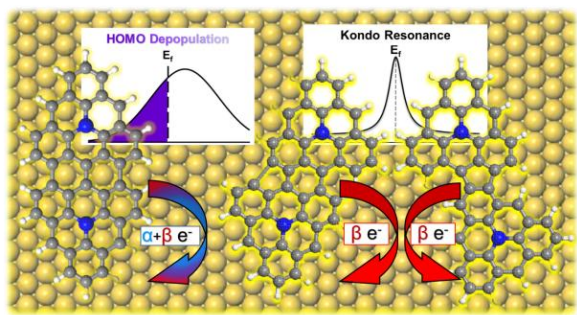

SYNOPSIS